\documentclass[11pt]{article}
\usepackage{graphicx}
\usepackage{array}
\usepackage{epsf} 
\usepackage{amsmath,color}
\usepackage{amssymb}
\usepackage{epsfig}
\usepackage{latexsym}
\usepackage{amsfonts}
\usepackage{dcolumn}
\usepackage{varioref}
\usepackage{ifthen}
\usepackage{graphicx}
\usepackage{amssymb,amsmath,amsthm}
\usepackage{slashed}
\begin{document}
\parindent 0mm 
\setlength{\parskip}{\baselineskip} 
\thispagestyle{empty}
\pagenumbering{arabic} 
\setcounter{page}{1}

\begin{center}
{\Large {\bf QCD determination of the leading order hadronic contribution to the muon g-2}}
\end{center}

\begin{center}
 C. A. Dominguez$^{(a)}$,
 K. Schilcher $^{(a),(b),(c)}$,  H. Spiesberger $^{(a)}$,$^{(c)}$ \\
\vspace{.2cm}

{\it $^{(a)}$ Centre for Theoretical and Mathematical Physics, 
	and Department of Physics, University of
	Cape Town, Rondebosch 7700, South Africa}\\
	
{\it $^{(b)}$ National Insitute for Theoretical Physics, 
Private Bag X1, Matieland 07602, South Africa}\\	

{\it $^{(c)}$ PRISMA Cluster of Excellence, 
Institut f\"{u}r Physik,
Johannes Gutenberg-Universit\"{a}t, 
D-55099 Mainz, Germany}\\

\end{center}

\begin{abstract}
\noindent
The leading order hadronic contribution to the muon magnetic moment anomaly, $a^{HAD}_\mu$, is determined entirely in the framework of QCD. The result in the light-quark sector, in units of $10^{-10}$, is $a^{HAD}_\mu|_{uds} =686 \pm 26$, and in the heavy-quark sector $a^{HAD}_\mu|_{c} =14.4 \pm 0.1$, and $a^{HAD}_\mu|_{b} =0.29 \pm 0.01$, resulting in $a^{HAD}_\mu = 701 \pm 26$. The main uncertainty is due to the current lattice QCD value of the first and second derivative of the electromagnetic current correlator at the origin. Expected improvement in the precision of these derivatives  may render this approach the most accurate and trustworthy determination of the leading order $a^{HAD}_\mu$.
\end{abstract}

\section{Introduction}
Some time ago  we proposed a novel method for determining the leading order hadronic contribution to the muon $g - 2$ entirely from theory, i.e. QCD \cite{SB1}-\cite{SB2}. The method is based on Cauchy's theorem in the complex squared energy $s$-plane, which provides a unique relation between the behaviour of current correlators on a circle of radius $|s|=s_0$ and their discontinuity across the real axis (see Fig.1). Here $s_0 \gtrsim 1 \;{\mbox{GeV}}^2$  is the threshold for perturbative QCD (PQCD) valid on the circle. This theorem states
\begin{equation}
\oint_C \Pi(s) \, ds = \sum_i \left[Residue\;\Pi(s)\; @ \; pole \right]_i \,, \label{Cauchy1}
\end{equation}
where $\Pi(s)$ is some QCD current correlator. After splitting the contributions on the circle and across the real axis, and introducing some arbitrary integration kernel, $K(s)$, it becomes
\begin{equation}
\frac{1}{2\,\pi\,i}\; \oint_{|s_0|} \Pi(s) \, K(s)\,ds \, + \,
\int_{s_{th}}^{s_0} \frac{1}{\pi} \; Im \, \Pi(s) \, K(s) \,ds =\, \sum_i \left[Residue \; \Pi(s)\, K(s) \,@ \; pole\right]_i \,.\label{Cauchy2}
\end{equation}
\begin{figure}
	[ht]
	\begin{center}
		\includegraphics[height=2.2in]{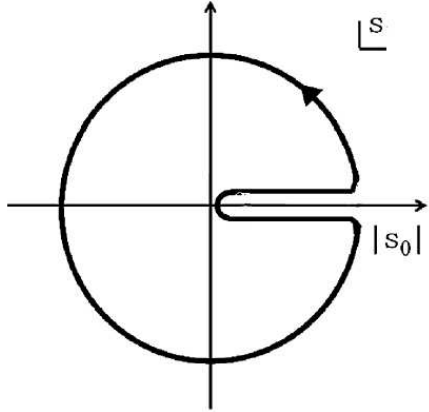}
		\caption{\footnotesize{The squared energy $s$-plane used in Cauchy's theorem, Eqs.(1)-(2).}}
	\end{center}
\end{figure}
This is called a QCD Finite Energy Sum Rule (FESR), which relates QCD information on the circle with e.g. hadronic physics on the real axis \cite{Shankar}. 
This duality between QCD and hadronic physics may be violated at low/intermediate energies \cite{Shankar}-\cite{DV}, 
$s_0 \lesssim 3 \, {\mbox{GeV}^2}$. However, in Section 2 we show that this effect is negligible in the present application.\\
One way of exploiting this fundamental relation is to determine the  low energy contribution encapsulated in the line-integral from information on the circle, provided  e.g. by perturbative QCD (PQCD), as well as from knowledge of the residues at the poles. The latter can be determined by Lattice QCD (LQCD). This leads to an entirely QCD determination of the anomaly which does not rely on experimental data e.g. from $e^+ e^-$
annihilation, or $\tau$- decay into hadrons.\\
The standard expression of the (lowest order) hadronic muon anomaly is given by \cite{review}
\begin{equation}
a_{\mu}^{HAD}=\;\frac{\alpha_{EM}^{2}}{3\,\pi^{2}}\,\int_{s_{th}=m_{\pi}^2}^{\infty
}\,\frac{ds}{s}\;K(s)\;R(s)\;,\label{AMUH1}%
\end{equation}
where $\alpha_{EM}$ is the electromagnetic coupling, $K(s)$ is a known integration kernel,  and the  $R$-ratio is 
\begin{equation}
R(s)=3\,\sum_{f}\,Q_{f}^{2}\left[  8\,\pi\,\mbox{Im}\,\Pi
(s)_{QCD}\right] \;, \label{RatioR}
\end{equation}
where $\Pi(s)_{QCD}$ is the QCD vector current correlator normalized as
\begin{equation}
\mbox{Im}\,\Pi(s)_{QCD}= \frac{1}{8\, \pi} \, \left[1 + \frac{\alpha_{s}}{\pi} + \cdots \right]\,. \label{ImPi}
\end{equation}
The integration kernel $K(s)$, at leading order, in Eq.(\ref{AMUH1}) is given by \cite{review}
\begin{equation}
K(s)=\int_{0}^{1}\,dx\,\frac{x^{2}(1-x)}{x^{2}+\frac{s}{m_{\mu}^{2}}
(1-x)}\;, \label{K}
\end{equation}
where $m_{\mu}$ is the muon mass. At leading order  one can split $a_\mu^{HAD}$  into the contributions from the three quark-mass regions labelled by the quark flavours $(u,d,s)$, $c$, and $b$, i.e.
\begin{equation}
a_{\mu}^{HAD}=a_{\mu}^{HAD}|_{uds}+a_{\mu}^{HAD}|_{c}+a_{\mu}^{HAD}
|_{b} \;. \label{amutotal}
\end{equation}
\begin{figure}[hb]
\begin{center}
\includegraphics[height=2.5 in] {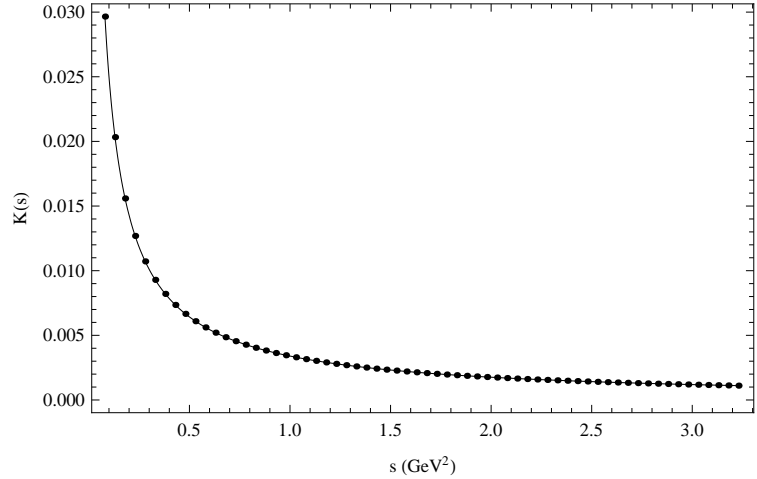}\caption{{\protect\footnotesize
{The exact kernel $K(s)$, Eq.(\ref{K}) (solid line), together with the fit kernel $K_1(s)$ (solid circles), Eq.(\ref{K1N}), in the
light-quark region. Relative difference is in the range $0-1$\%. The corresponding anomalies using all available experimental data are given in Eqs.(9)-(10).}}}
\label{figure2}
\end{center}
\end{figure}
In order to be able to determine each one of these contributions entirely from theory, thus making use of Cauchy's theorem, Eq.(\ref{Cauchy2}), it is necessary to substitute the original kernel $K(s)$ in Eq.(\ref{AMUH1}) by kernels possessing pole singularities. It turns out that given its shape, $K(s)$ can be easily substituted, with extreme accuracy, by such kernels in the three separate regions (uds),(c), and (b). Starting with the light-quark sector, the optimal substitute kernel is \cite{SB1}
\begin{equation}
K_{1}(s) = a_1\, s + a_2 \, s^{-1}  + a_3\, s^{-2} 
+  a_4 \,s^{-3} \;,\label{K1N}
\end{equation}
in the region $s_{th} \leq s \leq s_0 = (1.8 \;\mbox{GeV})^2$. The values of the coefficients are:  $a_1 =2.257\times 10^{-5}\,  {\mbox{GeV}^{-2}}$, 
$a_2 = 3.482\times 10^{-3} \, {\mbox{GeV}^{2}}$, $a_3 = - 1.467\times 10^{-4}\, {\mbox{GeV}^{4}}$, and $a_4 = 4.722\times 10^{-6}\, {\mbox{GeV}^{6}}$.  This is shown in Fig.2 (solid circles) together with the original kernel K(s) (solid line). The relative difference between $K_1(s)$ and $K(s)$ in this region lies in the range $0-1\,\%$.\\
A further estimate of this excellent accuracy can be obtained by using all available $e^+ e^-$ experimental data for $R(s)$ in this region together with the original kernel $K(s)$, and the substitute $K_1(s)$, to compute $a_{\mu}^{HAD}|_{uds}$.
\newpage
The results  for $s_0 = (1.8 \, {\mbox{GeV}})^2$ are (in units of $10^{-10}$)
\begin{equation}
a_{\mu}^{HAD}|_{uds} = 641.69 \;,\label{AMUUDS1}
\end{equation}
using the original kernel, $K(s)$, and
\begin{equation}
a_{\mu}^{HAD}|_{uds} = 641.16\;,\label{AMUUDS2}
\end{equation}
using  the fit kernel, $K_1(s)$, i.e. a difference of 0.08\%.

Proceeding to the charm-quark sector the fit kernel is   given by \cite{SB1}
\begin{equation}
K_{2}(s)= b_1 \,s^{-1}\;+\; b_2\, s^{-2}\;,\label{K2}
\end{equation}
where $b_{1}=0.003712\;\mbox{GeV}^{2}$, and $b_{2}=-0.0005122\;\mbox{GeV}^{4}$, in the range $M_{J/\psi}^{2} \leq s \leq s_2$, with $s_2 \simeq (5.0\,\mbox{GeV})^{2}$.
This function differs from the original  kernel $K(s)$ by less than $0.02\%$, thus providing an excellent fit. Finally, in the bottom-quark region, $M_{\Upsilon}^{2}\leq s\leq(12.0\,\mbox{GeV})^{2}$,
the optimal fit kernel is \cite{SB1}
\begin{equation}
K_{3} (s)=c_1\,s^{-1} + c_2\,s^{-2}\,, \label{K3}
\end{equation}
where $c_1 = 0.003719\,\mbox{GeV}^{2}$, and $c_2=-0.0007637\, \mbox{GeV}^{4}$. This kernel differs  from the exact kernel, $K(s)$, by less than 0.0005 \% in this range.\\
These three integration kernels $K_i(s)$ will be used to compute the contour and the line integrals in Eq.(\ref{Cauchy2}), up to the corresponding values of the Cauchy radius, $s_0$. Beyond these limits perturbative QCD (PQCD) can be safely used up to infinity to fully saturate the line integral in Eq.(\ref{AMUH1}), thus requiring only the original kernel $K(s)$.
Finally, the residues in Eq.(\ref{Cauchy2}) can be fully computed in PQCD in the charm- and bottom-quark sectors making use of the heavy-quark mass expansion at the origin, known up to the four-loop level. In the light-quark sector LQCD determinations of the first and second derivatives of the vector current correlator will be used to calculate the residues thus completing the theoretical calculation of the anomaly. As a complementary test in the light-quark region, the leading residue can also be determined from the electromagnetic radius of the pion, well known from data.

\section{QCD Determination of $a_\mu^{HAD}$}

The expression for the lowest order hadronic anomaly, Eq.(\ref{AMUH1}), can be recast as
\begin{equation}
a_\mu^{HAD} = 8 \alpha_{EM}^2 \sum_i Q_i^2 \left\{ \mbox{Res} \left[ \Pi_{i}(s) \frac{K_{i}(s)}{s}\right]_{s=0}  
- \frac{1}{2 \pi i} \oint_{|s|=s_0}\frac{ds}{s} \, K_{i}(s) \; \Pi_{i}(s) 
+ \int_{s_0}^{\infty} \, \frac{ds}{s} \; K(s) \; \frac{1}{\pi} \, \mbox{Im}\, \Pi_{i}(s)\right\}, \label{AMUL}
\end{equation}
where the index $i$ runs from one to three, covering the three sectors (uds), (c) and (b). Notice that the last term above involves the original kernel $K(s)$. Beginning with the charm-quark sector, the perturbative QCD heavy-quark Taylor series expansion of the correlator around the origin is
\begin{equation}
\Pi_{c}(s)|_{PQCD}=\frac{3}
{32\pi^{2}}\,Q_{c}^{2}\,\sum_{n\geq0}\bar{C}_{n}z^{n}\;,\label{Pic1}
\end{equation}
where $z=s/(4\bar{m}_{c}^{2})$. The  mass $\bar{m}_{c}\equiv\bar{m}_{c}(\mu)$ is the charm-quark mass in the $\overline{\text{MS}}$-scheme at a renormalization scale $\mu$. The coefficients $\bar{C}_{n}$ up to $n=30$ are known at three- and four-loop level \cite{boughezal2006a}-\cite{maier2010}. No coefficients $\bar{C} _{4}$ and higher contribute to the residue due to the s-dependence of $K_{2}(s)$.  Using as input $\mu=3\,\mbox{GeV}$, $\alpha_{s}^{(4)}(3\,\text{GeV})=0.2145(22)$ \cite{PDG}
and $\bar{m}_{c}(3\,\text{GeV})=0.986(10)\,\text{GeV}$ \cite{mc}, one finds
\begin{equation}
\Pi_{c}(s)=0.03604+0.001833\;s+0.00012335\;s^{2}
+0.000012472\;s^{3}+\mathcal{O}(s^{4})\;,\label{Pic2}
\end{equation}
where $s$ is expressed in $\mbox{GeV}^{2}$, and the coefficients have the appropriate units to render $\Pi_{c}(s)$ dimensionless. The residue in the charm-quark sector is
\begin{equation}
\text{Res}\left[  \Pi_{c}(s)|_{PQCD}\frac{K_{2}(s)}{s}\right]_{s=0}
=76.1(5)\,\times10^{-7}\;,\label{RESC}%
\end{equation}
where the error is due to the uncertainty in $\alpha_{s}$ and to the
truncation of PQCD. For the bottom quark sector the  residue is
\begin{equation}
\text{Res}\left[  \Pi_{b}(s)|_{PQCD}\frac{K_{3}(s)}{s}\right]_{s=0}
=6.3\,\times10^{-7}\;,\label{RESB}%
\end{equation}
where the error is negligible. Next, in order to calculate the contour
integral around the circle  we make use of PQCD, i.e.
\begin{equation}
\Pi_{\text{PQCD}}(s)=\sum_{n=0}^{\infty}\left(  \frac{\alpha_{s}(\mu^{2}%
)}{\pi}\right)  ^{n}\Pi^{(n)}(s)\;,\label{PIQCD}
\end{equation}
where 
\begin{equation}
\Pi^{(n)}(s)=\sum_{i=0}^{\infty
}\left(  \frac{\bar{m}^{2}}{s}\right)  ^{i}\Pi_{i}^{(n)}\;.\label{PIN}
\end{equation}
The complete analytical result in PQCD up to  $\mathcal{O}(\alpha_{s}^{2},(\bar{m}
^{2}/s)^{30})$ is given in \cite{Chetyrkin1997}-\cite{Baikov2009}, while $\Pi
_{2}^{(3)}$ is known up to a constant term \cite{Chetyrkin2000b}. This
constant term does not contribute to the contour integral due to the
s-dependence of $K_{2}(s)$. Finally, at five-loop level the
full logarithmic terms in $\Pi_{0}^{(4)}$ and $\Pi_{1}^{(4)}$ are known from \cite{Baikov2008} and \cite{Baikov2004}, respectively.\\
Putting all together, the  contour integrals using fixed order perturbation theory (FOPT) are
\begin{eqnarray}
\frac{1}{2 \pi i} \oint\frac{ds}{s} \; K_{n}(s) \; \Pi_{q}(s)|_{PQCD} \;  = \;\left\{ 
\begin{array}{lcl}
135.3 (6) \times 10^{-7} \\
\,\, \,20.3 (1) \times 10^{-7} \\
\, \,\,\,\,\,3.6 (2) \times 10^{-7}  \;,\label{OINT}
\end{array}\right.
\end{eqnarray}
for $n=1,2,3$ and $q=uds,c,b$, respectively. For $n=1$, i.e. the $(uds)$ sector, the result in contour improved perturbation theory (CIPT) is $135.6(6) \, \times\, 10 ^{-7}$, i.e. a 0.2\% difference with FOPT. Also for $n=1$, changing the PQCD threshold in the interval $s_0 = (1.8 - 2.0)^2\, \mbox{GeV}^2$ leads to a change of only  0.15\% in the final value of $a^{HAD}_{\mu}$. The BES Collaboration data in this region and beyond \cite{BES} agrees well with PQCD. The results for the line integral in Eq.(\ref{AMUL}) are
\begin{equation}
\int_{s_j}^{\infty}  \frac{ds}{s} K(s) \, \frac{1}{\pi} \, \mbox{Im}\, \Pi_{q}(s)|_{PQCD}=\left\{ 
\begin{array}{lcl}
151.8 (1)\times 10^{-7} \\
\,\,\,20.0 (4)\times 10^{-7} \\  
\,\,\,\,\,\,3.4 (2)\times 10^{-7} \label{INT}
\end{array}\right.
\end{equation}
with $s_j= (1.8, 5.0, 12.0) \,{\mbox{GeV}^2}$ for $q=uds,c,b$, respectively.  Substituting the results from
Eqs.(\ref{OINT}) and ({\ref{INT}) into Eq.(\ref{AMUL}), together with
the residues in the charm- and bottom-quark sectors, Eqs. (\ref{RESC})-(\ref{RESB}), the leading order $a_{\mu}^{HAD}$  are \cite{SB1}
\begin{equation}
a_{\mu}^{HAD}|_{c}=14.4(1)\times10^{-10}\;, \label{amuc}
\end{equation}
\begin{equation}
a_{\mu}^{HAD}|_{b}=0.29(1)\times10^{-10}\;. \label{amub}
\end{equation}
These results were fully confirmed later by  LQCD calculations yielding $a_{\mu}^{HAD}|_{c}=14.42(39) \times 10^{-10}$ from \cite{LQCD1}, and $a_{\mu}^{HAD}|_{b}=0.271(37)\times10^{-10}$ from \cite{LQCD2}.\\
The complete result for the anomaly can then be written as
\begin{equation}
a_{\mu}^{HAD}= \Bigg\{ \frac{16}{3}\alpha_{EM}^{2} \, \mbox{Res}\left[  \Pi_{uds}(s)\frac{K_{1}(s)}{s}\right]_{s=0} + 4.7(2) \times 10^{-10} \Bigg\}+ 14.7(1)\times 10^{-10}\;,\label{AMUHSF}
\end{equation}
where the term in curly brackets corresponds to the light quark sector, and the last term in the equation is the total charm- plus bottom-quark contribution. Using the quark-hadron duality violation model of \cite{Shifman} with parameters from \cite{DV}, we find this effect to be negligible in comparison with the first term inside curly brackets.\\
The first two derivatives of the light-quark current correlator at the origin have been calculated in LQCD \cite{Borsanyi} with the results
\begin{equation}
\frac{d}{dq^2} \, \Pi(s)_{uds}|_{s=0} = 0.07190 \pm 0.0025, \,\ {\mbox{GeV}^{-2}} \, \label{Pprime}
\end{equation}
\begin{equation}
\frac{d^2}{(dq^2)^2} \, \Pi(s)_{uds}|_{s=0} = 0.136 \pm 0.009, \, {\mbox{GeV}^{-4}} \,, \label{Ppprime}
\end{equation}
where these values correspond to our definition and normalization of the QCD correlator, Eqs.(4)-(5). Hence, the results of \cite{Borsanyi} must be multiplied by a factor $3/4$.
Notice that the second derivative is related to the third term in Eq.(8), which is an order of magnitude smaller, and of opposite sign than the second term, corresponding to the first derivative. Even though the second order pole residue is thus small compared with the first order one, it is not entirely negligible given the overall required accuracy. However, the contribution of the pole of third order can be safely neglected due to its factorial suppression, as well as the relative size of the overall coefficient in Eq.(8).
The  value of the residue in the light-quark sector becomes
\begin{equation}
\mbox{Res}\left[  \Pi_{uds}(s)\frac{K_{1}(s)}{s}\right]_{s=0} = (0.240 \pm 0.009)\times 10^{-3} \;,\label{RESuds}
\end{equation}
and the complete anomaly, Eq.(24), is
\begin{equation}
a_{\mu}^{HAD}= (701 \pm 26) \, \times 10^{-10}. \label{amuF}
\end{equation}
Further improvement in precision will require more accurate LQCD results for the derivatives of the electromagnetic current correlator.
\section{Electromagnetic pion radius contribution to $a_{\mu}^{HAD}|_{ud}$}
In closing, we discuss a procedure to relate the first derivative of the vector current correlator to the electromagnetic radius of the pion. This information is useful in the framework discussed in the previous section, Eqs. (24)-(25), i.e. to determine the residue of the current correlator in the light-quark region. The leading order hadronic representation of the vector current correlator in Eq.(4), normalized as in Eq.(5), can be written schematically as
\begin{equation}
\Pi_{\mu\nu}|_{HAD} (q^2) = i \int d^4x\,  e^{iqx}\, \Big\{  \langle 0| V_\mu(x)|\pi \pi \rangle \langle \pi \pi| V_\nu^\dagger(0)|0\rangle 
+ \langle 0| V_\mu(x)|\rho\rangle \langle \rho| V_\nu^\dagger(0)|0\rangle
 + \cdots\Big\} \,,\label{Vcorr}
\end{equation}
where a four-momentum integral, and a sum over polarizations (for the $\rho$-meson) is implicit, and the neglected terms are the multiple-pion matrix elements. The first term in the integral above is loop-suppressed with respect to the Born, single rho-meson contribution. A straightforward calculation of the  second term in Eq.(29) gives
\begin{equation}
\Pi_{\mu\nu}|_{HAD} (q^2) = (q_\mu q_\nu - q^2 g_{\mu\nu}) \Pi_0(q^2) \,,
\label{PiHAD}
\end{equation}
with
\begin{equation}
\frac{d}{dq^2} \, \Pi_{0}(s)|_{s=0} = \frac{1}{f_\rho^2} \, \frac{1}{6} \langle r^2_{\pi}\rangle = 0.0764 \pm 0.0015 \, {\mbox{GeV}^{-2}},\label{radius}
\end{equation}
where $f_\rho = 4.96 \pm 0.02$ \cite{PDG}, and $\langle r^2_{\pi}\rangle = 0.439 \pm 0.008 \;{\mbox{fm}^2}$ \cite{rpi}. Using this result would give $a_{\mu}^{HAD}= (775 \pm 14) \, \times\, 10^{-10}$. After subtracting a potential 5\% contribution from a second derivative gives $a_{\mu}^{HAD}= (736 \pm 14) \, \times\, 10^{-10}$. These results, while involving a few approximations/assumptions, provide further support for the value obtained in Eq.(28) using current LQCD information on the first two derivatives of the vector current correlator at the origin. The latter should be known in future from LQCD with much improved precision.
\section{Summary}
In this paper we  made use of a novel method, first proposed in \cite{SB1}, to determine the leading hadronic contribution to the muon magnetic moment anomaly, $g-2$, entirely from theory. Given that this quantity has been exploited intensively as the culprit for Physics beyond the Standard Model, it is imperative to determine it in the framework of our current strong interaction theory, i.e. QCD.
The essential tool to perform this task is Cauchy's theorem in the complex squared-energy plane (Fig.1), proposed long ago \cite{Shankar} to relate QCD to hadronic Physics. Given the absence of singularities in the complex squared-energy $s$-plane, except for a discontinuity across the real $s$-axis (due to hadronic poles and resonances), one can relate QCD information on a circle of radius $|s_0|$ to hadronic information on the real axis. Here, the radius $|s_0|$ is chosen large enough for perturbative QCD to be valid on the circle, as well as large enough to cover relevant hadronic contributions. In order to exploit Cauchy's theorem to the fullest it is necessary to replace the integration kernel $K(s)$ entering the anomaly, Eq.(3), by meromorphic kernels. These require information on the first few derivatives of the electromagnetic correlator at the origin, which is being determined by LQCD \cite{LQCD1}-\cite{LQCD3}.
These substitute kernels in the light-, charm-, and bottom-quark sectors are essentially indistinguishable from the original, as witnessed by the negligible differences of $0 - 1 \%$ in the $(uds)$-region, $0-0.02 \%$  in the charm-sector, and $0-0.0005 \%$ for bottom. Furthermore, a crucial test was performed in the dominant light-quark sector by using all the available $e^+ -e^-$ data together with the original kernel to compute $a^{HAD}_\mu$, and compare with the result from using the substitute kernel $K_1(s)$. Fully supporting results are shown in Eqs.(9)-(10), respectively.\\
Results from this approach in the charm- and bottom-quark sectors, Eqs.(\ref{amuc})-(\ref{amub}), are in full agreement with LQCD determinations \cite{LQCD1}-\cite{LQCD2}, thus providing additional validation of our method. Nevertheless, the bulk of the contribution to the anomaly arises from the light-quark sector, as indicated in Eq.(\ref{AMUHSF}). The great challenge is for LQCD to provide values for the derivatives of the electromagnetic current correlator at the origin with improved accuracy. A hint on a potential outcome is provided by the result for the anomaly obtained using the experimental value of the electromagnetic radius of the pion in Section 3.
\section{Acknowledgements}
This work was supported in part by the National Research Foundation (South Africa), the Deutsche Forschungsgemeinschaft (Germany), and the National Institute for Theoretical Physics (South Africa).
One of us (CAD) wishes to thank the organizers of the conference {\it Les Rencontres de Physique de la Vallee d'Aoste, 2017}, where a preliminary version of these results was presented. Helpful correspondence with Laurent Lellouche is greatly appreciated, as well as discussions with Gilberto Colangelo, Mario Greco,  Alvarao de Rujula, Hanno Horch, and Hartmut Wittig.


\begin{thebibliography}{0}    
\bibitem{SB1} S. Bodenstein, C. A. Dominguez, and K. Schilcher, Phys. Rev. D {\bf 85}, 014029 (2012).
\bibitem{SB2} C. A. Dominguez, K. Schilcher, and H. Spiesberger, Mod. Phys. Lett. A {\bf 31}, 1630035-1 (2016).
\bibitem{Shankar} R. Shankar, Phys. Rev. D {\bf 15}, 755 (1977).    
\bibitem{Shifman} M. A. Shifman, Prog. Theor. Phys. Suppl. {\bf 131} (1998) 1.
\bibitem{DV}A. Pich and A. Rodriguez-Sanchez, Mod. Phys. Lett. A {\bf 31}, 1630032-1 (2016) ; M. Gonzalez Alonso, A. Pich, and A. Rodriguez-Sanchez, Phys. Rev. D {\bf 94}, 014017 (2016).
\bibitem {review}J. P. Miller, E. de Rafael, and B. Lee Roberts, Rep. Prog.
Phys. \textbf{70}, 795 (2007); F. Jegerlehner, and A. Nyffeler, Phys. Rep.
\textbf{477}, 1 (2009),  and references therein.
\bibitem {boughezal2006a}R. Boughezal, M. Czakon, and T. Schutzmeier, Phys. Rev. D \textbf{74}, 074006 (2006).
\bibitem {maier2008a}A. Maier, P. Maierh\"{o}fer, and P. Marquard, Nucl. Phys. B \textbf{797}, 218 (2008); Phys. Lett. B \textbf{669}, 88 (2008).
\bibitem {chetyrkin2006}K. G. Chetyrkin, J. H. K\"{u}hn, and C. Sturm, Eur.
Phys. J. C \textbf{48}, 107 (2006).
\bibitem {maier2010}A. Maier \textit{et al.}, Nucl. Phys. B \textbf{824}, 1 (2010).
\bibitem{PDG} K. Nakamura \textit{et al.}, Particle Data Group, J. Phys. G \textbf{37}, 075021 (2010).
\bibitem{mc} K. G. Chetyrkin  \textit{et al.}, Phys. Rev. D {\bf 80}, 074010 (2009); S. Bodenstein \textit{et al.}, Phys. Rev. D {\bf 82}, 114013 (2010); {\bf 83}, 074014 (2011).
\bibitem{Chetyrkin1997} K. G. Chetyrkin \textit{et al.}, Nucl. Phys. B \textbf{503}, 339 (1997).
\bibitem{MAIER2} A. Maier, and P. Marquard, arXiv:1110.558.
\bibitem {Baikov2009}P. A. Baikov, K. G. Chetyrkin, and J. H. K\"{u}hn, Nucl. Phys. B (Proc. Suppl.) \textbf{189}, 49 (2009).
\bibitem {Chetyrkin2000b}K. G. Chetyrkin, R. Harlander, and J. H. K\"{u}hn,
Nucl. Phys. B \textbf{586}, 56 (2000).
\bibitem {Baikov2008}P. A. Baikov, K. G. Chetyrkin, and J. H. K\"{u}hn, Phys. Rev. Lett. \textbf{101}, 012002 (2008).
\bibitem {Baikov2004}P. A. Baikov, K. G. Chetyrkin, and J. H. K\"{u}hn, Nucl. Phys. B (Proc. Suppl.) \textbf{135}, 243 (2004).
\bibitem{BES} J. Z. Bai {\it et al.}, BES Coll., Phys. Rev. Lett. {\bf 88}, 101802 (2002);
M. Ablikim {\it et al.}, BES Coll., Phys. Lett. B {\bf 677}, 239 (2009).
\bibitem{LQCD1} B. Chakraborty {\it {et  al.}}, Phys. Rev. D \textbf {89}, 114501 (2014);
 J. Koponen {\it {et  al.}}, arXiv:1411.0569 (2014). 
\bibitem{LQCD2} B. Colquohoun {\it {et  al.}}, Phys. Rev. D \textbf{91}, 074514 (2015). 
\bibitem{Borsanyi} Sz. Borsanyi {\it et al.}, arXiv: 1612.02364.
\bibitem{LQCD3} F. Burger {\it {et  al.}}, J. High Ener. Phys. 
\textbf{1402}, 099 (2014).
\bibitem{rpi} S. R. Amendolia {\it et al.}, Nucl. Phys. B {\bf 277}, 168 (1986).
\end{thebibliography}
\end{document}